\documentclass{aa}
\usepackage{graphics,latexsym,amssymb,times,psfig}

\def\gsim{ \lower .75ex \hbox{$\sim$} \llap{\raise .27ex \hbox{$>$}} } 
\def\lsim{ \lower .75ex\hbox{$\sim$} \llap{\raise .27ex \hbox{$<$}} } 

\begin{document}

\title{Emission lines in GRBs constrain the total energy reservoir}

\author{Gabriele Ghisellini \inst{1}, Davide Lazzati \inst{2},
Elena Rossi \inst{2} and Martin J. Rees \inst{2}}

\offprints{G. Ghisellini; gabriele@merate.mi.astro.it}
\institute{
Osservatorio Astronomico di Brera, via Bianchi 46, I--23807 Merate, Italy;
\and Institute of Astronomy, Madingley Road, CB3 0HA Cambridge, UK}

\date{Received 2001}
 
\titlerunning{Lines in GRBs and the energy budget}
\authorrunning{G. Ghisellini, D. Lazzati, E. Rossi \& M. Rees}

\abstract{
The emission features observed in the X--ray afterglow of Gamma Ray
Bursts are extremely powerful. Since they last at least for several
hours, they imply energies of the order of $10^{49}$ ergs.  This in
turn implies that the energy contained in the illuminating continuum
thought to be responsible of the line production must exceed $10^{51}$
ergs.  This is a strong lower limit to the energy reservoir of Gamma
Ray Bursts, which is independent of collimation and beaming, and bears
important consequences on the possible collimation of the fireball
radiation and the density of the medium surrounding the burst.
}
\maketitle

\section{Introduction}

The discovery that Gamma Ray Bursts (GRBs) are cosmological implies
large luminosities, but the exact value of the radiated luminosities
and the kinetic power of the fireball originating the emission are
still uncertain by a large factor, since we do not know if the
emission is isotropic or if it is instead collimated in a cone.  The
main tool to estimate the degree of collimation of the emission has
been, so far, the presence of an achromatic break in the lightcurves
of the afterglows.  This is interpreted as due to the deceleration of
the fireball, whose bulk Lorentz factor $\Gamma$ becomes smaller than
the inverse of the jet opening angle $\theta_{\rm j}$ (see e.g. Rhoads
1999).  This allows to estimate $\theta_{\rm j}$ and to obtain the
``true" values of the emitted energy.  Most notably, in this respect,
is the finding of Frail et al.  (2001) who found a remarkable
``clustering" of the fireball energy, once they are corrected by the
estimate of their degree of collimation.  Since the total power and
energy are obviously the main parameters for the construction of any
model, the importance of these estimates is obvious.  These in turn
are based on a number of assumptions, such as the density of the
matter surrounding the burst site, which is responsible for the
deceleration of the fireball, and on the key observation of the
presence of an achromatic break in the light curve of the afterglow.

Independent estimates of the ``true" energy, or even limits on it, are
called for.  In this paper we point out that the observed emission
features observed in the X--ray afterglow spectra of several bursts
can indeed put a firm lower limit to the emitted luminosity of GRBs.
This limit is at the same time simple, independent of the degree of
collimation of the burst radiation, and independent of the density of
the surrounding medium.

\section{Energy of the observed emission features}

Table 1 reports the main information obtained for the five bursts for
which emission features have been detected so far.  For GRB 970828 the
redshift listed in Table 1 is uncertain since this burst did not have
a standard optical or radio afterglow and therefore the identification
of the host galaxy should be regarded as tentative.  For GRB 000214
the value of redshift listed in Table 1 is even more uncertain, since
it corresponds to assume that the feature observed in X--rays is a
iron 6.97 keV emission line.

The significance of each of the line detections, so far, 
is at the level of only 3--4$\sigma$, with the iron line of
GRB 991216 (observed by Chandra) being the best evidence
(see references listed in Tab. 1).
Thus their reality is still under debate, and the issue will likely
be settled only with future observations,
even if the fact that they have been ``observed" so far by 
the $Beppo$SAX, ASCA, Chandra and XMM--Newton satellites
brings some weight to their reality.

The line luminosities listed in Table 1 are derived under the
assumption of isotropic emission (but see below); the start time
$t_{\rm s}$ corresponds to the beginning of the X--ray observation
which led to the discovery of the emission feature, while the end time
$t_{\rm e}$ corresponds to the time for which either the line was not
detected any longer or to the ending of the observations.  We give, in
Table 1, two values of the total line energy: the first assumes that
the line existed only for the time interval $(t_{\rm e}-t_{\rm s})$
(``short lived" line), while the second assumes that the line remained
constant in flux for $(t_{\rm e}-0)$ (``long lived" line).
Indications in favor of a line of constant flux come from GRB 000214
(Antonelli et al. 2000), while for GRB 970508 and GRB 011211 the
emission lines became undetectable before the end of the X--ray
observation.  
In all cases, despite the range of start times
of X--ray observations in different bursts, the line was always
visible in the first part of the observations (i.e. there is no
indication of line fluxes increasing in time).
\begin{table*}
\begin{center}
\begin{tabular}{|l|lllllllllll|}
\hline
&&&&&&&&&&&\\ 
GRB     &$z_{\rm opt}$ &$F^{\rm line}_{-14}$   &$\epsilon^{\rm line}$ &$t_{\rm s}$--$t_{\rm e}$ &$L^{\rm iso}_{44}$   
        &($E^{\rm iso}_{49})^a$ &($E^{\rm iso}_{49})^b$ &Ref  &$\theta^{\rm F01}_{\rm j}$ &$E^{\rm F01}_{\gamma, 51}$ 
        &$E_{\gamma,51}$\\
        &              &erg s$^{-1}$cm$^{-2}$  &keV                   &h     &erg s$^{-1}$  &erg       &erg 
        &                               &deg                    &erg  &erg  \\
\hline 
970508 &0.835 &$30\pm 10$    &$3.4\pm0.3$  &6--16    &$12\pm4$     &$2.25\pm 0.8$   &$3.6\pm1.3$   &P99 &16.7 &0.23 &11--18 \\
970828 &0.958?&$15\pm8$      &$5\pm0.25$   &32--38   &$8.1\pm4.3$  &$0.9\pm0.4$     &$5\pm3$       &Y99 &4.1  &0.57 &4.5--25 \\ 
991216 &1.02  &$17\pm5$      &$3.5\pm0.06$ &37--40   &$11\pm3$     &$0.6\pm0.2$     &$7.7\pm2.3$   &P00 &2.9  &0.69 &3--38 \\ 
000214 &0.46? &$6.7\pm2.2$   &$4.7\pm0.2$  &12--41   &$0.6\pm 0.2$ &$0.4\pm0.14$    &$0.6\pm0.3$   &A00 &---  &---  &2--3 \\ 
011211 &2.14  &$0.8\pm0.5$   &Mg           &11--12.4 &$3\pm2$      &$0.05\pm 0.04$  &$0.4\pm0.3$   &R02 &---  &---  & \\
011211 &2.14  &$1.1\pm0.3$   &Si           &11--12.4 &$4.3\pm1.2$  &$0.07\pm 0.02$  &$0.6\pm0.2$   &R02 &---  &---  & \\
011211 &2.14  &$1.\pm0.3$    &S            &11--12.4 &$3.9\pm1.1$  &$0.06\pm 0.02$  &$0.55\pm0.16$ &R02 &---  &---  &  \\
011211 &2.14  &$0.7\pm0.25$  &Ar           &11--12.4 &$2.6\pm1$    &$0.04\pm 0.016$ &$0.4\pm0.14$  &R02 &---  &---  & \\
011211 &2.14  &$0.44\pm0.22$ &Ca           &11--12.4 &$1.7\pm0.9$  &$0.03\pm 0.015$ &$0.25\pm0.12$ &R02 &---  &---  & \\
011211 &2.14  &$4.\pm1.6$    &Sum          &11--12.4 &$15.6\pm6.3$ &$0.25\pm 0.1$   &$2.2\pm0.9$   &R02 &---  &---  &0.5--4.4 \\
\hline
\end{tabular}
\caption{Luminosities and energies are calculated assuming 
$H_0=65$ km s$^{-1}$ Mpc$^{-1}$, $\Omega_{\Lambda}=0.7$ and
$\Omega_{\rm m}=0.3$, and assuming that the line emission is
isotropic. We use the notation $Q\equiv 10^x Q_x$. 
The redshift of GRB 000214 is calculated assuming that the
observed line is from H--like iron.  The time $t_{\rm s}$ is the start
time of the X--ray afterglow observations; $t_{\rm e}$ is the time
until which the line was visible.  $(E^{\rm iso}_{49})^a$ corresponds
to assuming that the line was constant in flux (and emitted
isotropically) in the time interval $(t_e-t_s)$; $(E^{\rm iso}_{49})^b$ 
corresponds to assuming a constant line in the time
interval $(t_e-0)$.  These values take into account the cosmological
time dilation factor $(1+z)$.  The last three columns report the jet
opening angle and the corrected energy emitted in $\gamma$--rays as
calculated by Frail et al. (2001) and our estimated range of $E_\gamma$
using $\eta_{\rm line}=0.02$ for all bursts but GRB 011211, 
for which we used $\eta_{\rm line}=0.05$.
References: 
A00: Antonelli et al. 2000; 
P99: Piro et al. 1999;
P00: Piro et al. 2000; 
R02: Reeves et al. 2002; 
Y99: Yoshida et al. 1999.}
\end{center}
\end{table*}
\subsection{Line photon collimation}
\begin{figure}
\psfig{file=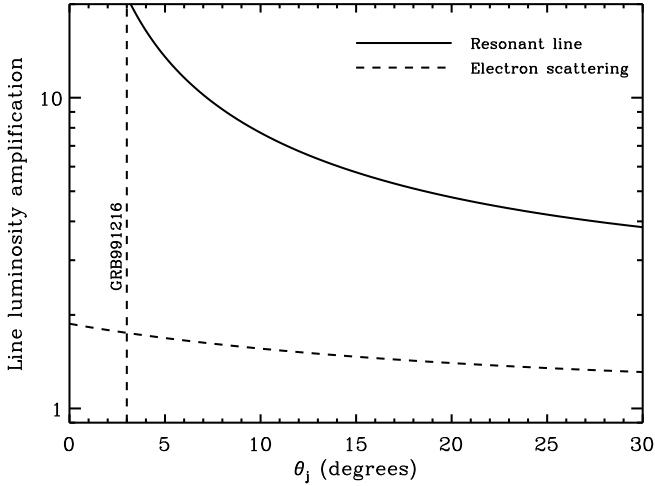,width=0.48\textwidth}
\caption{{Amplification of the line luminosity in a funnel after 
line photons are reprocessed by scattering off the funnel wall.}
\label{fig:beam}}
\end{figure}

Even for line photons produced isotropically, it is possible that
multiple scatterings off the walls of a funnel produce an effective
collimation of the emission, enhancing its observed flux (McLaughlin
et al. 2002) by the factor $4\pi/\Omega_{\rm line}$ with respect to
isotropic emission.  Therefore the line photons may be ``channeled"
into a cone of opening angle comparable to the funnel opening angle.
There are however two effects limiting the collimation efficiency.
First, photons scattered by the wall of the funnel close to its
opening can escape in a direction different from that of the line of
sight.  These photons will be lost and not redirected towards the
observer.  Second, for each scattering the line photon energy changes
if the scattering is not completely elastic, resulting in the smearing
of the line. If the scattering opacity is mainly due to free electrons
at a temperature of several keV, the line width after several
scatterings (Pozdniakov et al. 1983) is $\sigma_\epsilon/\epsilon =
(2N_{\rm{sc}}kT/m_ec^2)^{1/2}$. The detected lines have widths
$\sigma_\epsilon/\epsilon\sim0.1$, from which we can derive
$N_{\rm{sc}}\sim3$.

Let us now consider a funnel of height $R$ and opening angle
$\theta_{\rm j}$. A photon scattered at a distance $Rz$
($0\le{}z\le1$) from the funnel opening will have a probability
$P_{\rm{obs}}=(1-\cos\theta_{\rm j})/2$ of being scattered towards the
observer, and a probability
$P_{\rm{esc}}(z,\theta)=(1-\cos[{\rm{atan}}(\sin\theta/z)])/2$ of
being scattered in any direction outside the funnel cone.  Assuming
that the funnel walls radiate and scatter the line uniformly, the
total probability for a photon to exit the funnel is:
\begin{equation}
P_{\rm{esc}}(\theta)=2\int_0^1(1-z)P_{\rm{esc}}(z,\theta)\,dz
\label{eq:pesc}
\end{equation}
In addition we must consider that half of the photons will be
scattered in the forward direction, i.e. deeper into the funnel walls 
rather than towards the open space. 
These photons will be lost if the main
scattering opacity is provided by free electrons. On the other hand,
if the considered line is resonant, these photons may be rescattered
by ions while still at a small Thomson depth in the walls and
redirected towards the open space. The total probability of a photon
to be lost is then $P_{\rm{lost}}=0.5+P_{\rm{esc}}(\theta)$ for
electron scattering and $P_{\rm{lost}}=P_{\rm{esc}}(\theta)$ for a
resonant line.

\begin{figure*}
\centerline{\psfig{file=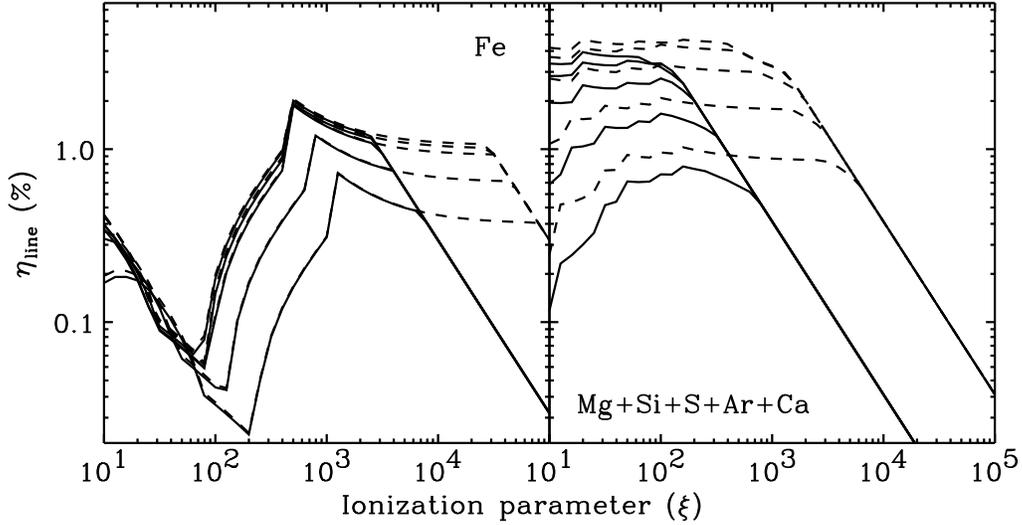,width=.75\textwidth}}
\caption{{Efficiency of conversion of continuum [1--30 keV] luminosity
into $K_\alpha$ line luminosity as a function of the ionization
parameter $\xi\equiv 4\pi\,F_{[1-30]{\rm keV}}/n$. The continuum is
assumed to be a power--law $F(\nu)\propto\nu^{-\alpha}$ within the
[1--30 keV] energy range. The left panel shows iron $K_\alpha$
emission, while the right panel shows the sum of Mg, Si, S, Ar and Ca
$K_\alpha$ lines.  Different solid lines show the efficiency for
$\alpha=0$, 0.5, 1, 1.5 and 2. Dashed lines are obtained with the same
parameter set as for solid lines, but with ten times solar iron
abundance. The iron peak efficiency $\eta_{\rm{line}}\approx 0.02$ is
obtained for $0<\alpha<1$. The soft X--ray lines peak efficiency
$\eta_{\rm{line}}\approx 0.05$ is obtained for $\alpha>1.5$.}
\label{fig:fe}}
\end{figure*}

After $N_{\rm{sc}}$ scatterings, the total number of photons emitted
in the direction of the observer will be enhanced by a factor
\begin{equation}
{\cal F}(\theta)=\sum_{n=0}^{N_{\rm{sc}}} 
\left[1-P_{\rm{lost}}(\theta)\right]^n \;;
\quad \lim_{N_{\rm{sc}}\to\infty} {\cal F}(\theta)=
{{1}\over{P_{\rm{lost}}(\theta)}}
\label{eq:spesc}
\end{equation}
Fig.~\ref{fig:beam} shows the results of the integral in Eq. 1 as a
function of the funnel opening angle $\theta_{\rm j}$. The solid lines
show the rather important amplification that a resonant line can
achieve (with infinite number of scatterings), while the dashed line
shows that, if electron scattering is important, the amplification of
the line can be at most a factor of two (the 3 scattering
amplification is shown, being very similar to the infinite one).

\section{Energy of the photoionizing continuum}

The efficiency of conversion of the X--ray ionizing continuum into
$K_\alpha$ line photons has been recently investigated by Lazzati,
Ramirez--Ruiz \& Rees (2002).  They assume that the line is produced
by reflection off an optically thick slab of material, illuminated by a
power--law continuum.  This is the more efficient way to reprocess
continuum photons into $K_\alpha$ lines.

Fig.~\ref{fig:fe} shows the result of the efficiency calculations for
iron and for the sum of the light elements Mg, Si, S, Ar and Ca.  We
explore the effects of different ionizing spectral shapes and
metallicity.  The reprocessing efficiency for iron cannot be larger
than $2\%$, while the combined light elements, for very small
ionization parameters, can reprocess up to $5\%$ of the continuum into
soft X-ray narrow lines.

It is important to note that the iron line is resonant for
$10 \lsim \xi \lsim 300$ and for $\xi > 10^4$. In these ranges the
efficiency of line production is much smaller due to the Auger
disruption of the line (see Ross, Fabian \& Brandt 1996). This effect
compensate for the possible line amplification due to beaming and we
will therefore concentrate on non--resonant lines in the following,
bearing in mind that similar results also apply to the resonant case.

\section{Limits on the total energy reservoir}

The energy contained in the emission lines sets a lower limit to both
the total radiated energy and to the total energy reservoir of the
fireball.  In fact the line energy is a fraction $\eta_{\rm line}$ of
the illuminating ionizing [1--30 keV] continuum, which is in turn a
fraction $\eta_x$ of the energy emitted in $\gamma$--rays during the
prompt emission.  We take into account that the line photons might be
collimated, resulting in an amplification factor $4 \pi /\Omega_{\rm
line}$ with respect to to the isotropic case.  Considering the
efficiency for iron lines (the most commonly detected) the limit to
the energy radiated in $\gamma$--rays then reads:
\begin{eqnarray}
E_\gamma\, &\ge&\,2\, {E^{\rm iso}_{\rm line}  \over  \eta_{x} 
\eta_{\rm line} }\, {\Omega_{\rm line} \over 4 \pi} 
\nonumber\\
&=&
500 E^{\rm iso}_{\rm line} \,\left({0.1 \over \eta_x}\right)
\left({0.02 \over \eta_{\rm line}}\right) 
\left({\Omega_{\rm line}/4\pi \over 0.5}\right)
\label{eq:e}
\end{eqnarray}
where the factor 2 in the first line of Eq.~\ref{eq:e} corresponds to
consider two--sided jets with the line emitting material visible on
only one side.  To find the total energy reservoir in the form of
kinetic energy of the fireball we should know the fraction $\eta_{\gamma}$
of it which is radiated.  
The total fireball energy is then $E=E_\gamma/\eta_\gamma$.
A value $\eta_\gamma\sim 0.2$ assumes that the conversion of bulk
kinetic energy into radiation is very efficient, and it is the same
value used by Frail et al. (2001).  

$\eta_x\sim 0.05$--0.1 is the value commonly observed during the
prompt emission of most bursts: it corresponds to a spectrum
$F(\nu)\propto \nu^0$ up to 300 keV.  This implicitly assumes that
$\eta_x$ is the same inside the entire collimation cone.  However, we
could have different scenarii, in which the emission at large angles
(which is probably the one illuminating the line emitting material) is
more X--ray rich than along the jet axis:

\begin{itemize}
\item
In the model of M\'esz\'aros \& Rees (2001), the line
emitting material is located in dense clumps inside an hypernova
envelope, illuminated by a non--thermal continuum corresponding to the
side expansion of the emerging jet.  This continuum could peak at the
``right" X--ray energies, but since it does not share the time--decay
properties of a ``standard" afterglow, its non--detection implies that
its total energy (i.e. its luminosity integrated for the emission
time) is less than the total energy contained in the X--ray afterglow.
This implies that also in this case $\eta_x<0.1$.

\item There could be a $\theta$--dependence of the bulk Lorentz 
factor and the fireball energy, as in the model by Rossi, Lazzati \&
Rees (2002). In this case it is possible that also the emission
properties can change with $\theta$, making $\eta_x$ a function of
$\theta$, possibly reaching $\eta_x$=1 at large angles.  In this model
$\theta_{\rm j}$ is assumed to be quite large ($>25^\circ$), 
implying no or weak collimation of the line photons
(i.e. $4\pi/\Omega_{\rm line} \to 1$).
Furthermore, the energy emitted
at large angles is in this model only a fraction of the total, making
the lower limit derived by Eq.~\ref{eq:e} approximately valid also in
this case.

\end{itemize}

\section{Discussion}

The most severe lower limit to the total energy reservoir is obtained
for GRB 991216 if its iron line ``lived" for 40 hours (i.e from soon
after the trigger until the end of the Chandra observation).  In this
case we obtain $E_{\rm line}=(7.7\pm 2.3)\times 10^{49}$ erg
corresponding by Eq.~\ref{eq:e} to the lower limit 
$E_\gamma\gsim 3.8\times 10^{52}$ erg for the energy radiated in $\gamma$--rays.  
Instead, if the line was ``short lived" and existed only during the 
3 hours of the Chandra exposure, we obtain 
$E_{\rm line} =(5.8\pm 2) \times 10^{48}$ erg, 
corresponding to $E_\gamma\gsim 2.8\times 10^{51}$ erg.
We then obtain a value which is a factor 4 to 50 greater than what
estimated by Frail et al. (2001),
which derived a jet opening angle of $\theta_{\rm j}\sim 2.9^\circ$ 
from the break in the lightcurve 1.2 days after the trigger, and
assuming an interstellar medium of density $n=0.1$ cm$^{-3}$.
The choice of this somewhat small value for the the circumburst 
density is in agreement with values derived by modelling  broad--band 
afterglow lightcurves (Panaitescu \& Kumar 2001).
Since the calculated fireball energy depends on $n^{1/4}$, the lower
limit estimated here becomes consistent with achromatic breaks being
due to collimated afterglow if the circum--burst density 
is $n> 25$ cm$^{-3}$ (short lived line) or 
$n> 6\times 10^5$ cm$^{-3}$ (long lived line). 
Since the total fireball power depends
from the square of the opening angle of the jet, we derive in this case
$\theta_{\rm j}>5.8^\circ$ or $\theta_{\rm j}>20^\circ$ for the
short or long lived line, respectively.

For GRB 970508 we derive $E_\gamma\gsim 10^{52}$ erg,
a factor $\sim$50 above the estimate of Frail et al. (2001) for the
same burst. 
This is particularly puzzling, since for this burst the radio data
(the source was followed for 450 days) can be used to calculate the
energy reservoir in an independent way, which yields $E\sim 5\times
10^{50}$ erg (Frail, Waxman \& Kulkarni, 2000), consistent with the
value of Frail et al. (2001), but rather inconsistent with the iron
line data.  
Wijers \& Galama (2000), fitting the afterglow spectrum of this burst 
(from radio to X--rays), derived a somewhat larger value
$(dE/d\Omega)\sim 3\times 10^{52}$ ergs$/4\pi$ sr, corresponding, for
$\theta_{\rm j}=16.7$ degrees, to $E\sim 1.6\times 10^{52}$ erg (two jets).  
Again, an increase of the circumburst density to 
$n\sim 6\times 10^5$ cm$^{-3}$ makes our limit consistent with the value of
the fireball energy derived by the presence of the achromatic break in
the afterglow lightcurve.

In GRB 011211 we only see emission line for elements lighter than iron
(Reeves et al. 2002; see Lazzati, Ramirez--Ruiz \& Rees 2002 for discussion), 
and correspondingly used $\eta_{\rm line} =0.05$ to derive the lower limits 
$E_\gamma\gsim 5\times 10^{50}$ erg and 
$E_\gamma\gsim 4.4\times 10^{51}$ erg for a short or long lived line, 
respectively.

\section{Conclusions}

{\it If} real, the line emission features observed in the X--ray
afterglow of GRBs are very luminous, and pose strong limits to the
total energetics of GRBs.  Since these limits are almost unaffected by
the abundance of metals, emission mechanisms and collimation of the
illuminating continuum, we consider them to be quite robust.  If we
want to reconcile these limits with the estimates of the jet opening
angle and total energy reservoir derived through the achromatic breaks
of the afterglow lightcurve, we are led to consider much larger
densities of the material surrounding the bursts, in contrast
with the results from broad band spectral fitting (e.g. Panaitescu \&
Kumar 2001).

We conclude by stressing the importance of the X--ray follow--up
observations, such as the one already possible with Chandra and
XMM--Newton, and the future one which can be performed by the Swift
satellite.  Observing emission features in the X--ray spectra can
immediately set powerful limits to the total energetics, while
observations of the time duration of the line (and its possible
variability behavior) can shed information on the illuminator and on
the geometry of the line emitting material.

\begin{acknowledgements}
GG, DL and ER thank the MIUR for the n. 2001-02-43 grant. 
ER thanks the Isaac Newton and PPARC 
scholarships for financial support. 
\end{acknowledgements}

\end{document}